# Symmetry-breaking bifurcations and ghost states in the fractional nonlinear Schrödinger equation with a PT-symmetric potential

Pengfei Li,[1,2,*] Boris A. Malomed, [3,4] and Dumitru Mihalache[5]

[1]*Department of Physics, Taiyuan Normal University, Jinzhong, 030619, China*
[2]*Institute of Computational and Applied Physics, Taiyuan Normal University, Jinzhong, 030619, China*
[3]*Department of Physical Electronics, School of Electrical Engineering, Faculty of Engineering, and Center for Light-Matter Interaction, Tel Aviv University, Tel Aviv 69978, Israel*
[4]*Instituto de Alta Investigación, Universidad de Tarapacá, Casilla 7D, Arica, Chile*
[5]*Horia Hulubei National Institute of Physics and Nuclear Engineering, Magurele, Bucharest RO-077125, Romania*
*Corresponding author: lpf281888@gmail.com*

**Abstract** We report symmetry-breaking and restoring bifurcations of solitons in a fractional Schrödinger equation with the cubic or cubic-quintic (CQ) nonlinearity and a parity-time (*PT*)-symmetric potential, which may be realized in optical cavities. Solitons are destabilized at the bifurcation point, and, in the case of the CQ nonlinearity, the stability is restored by an inverse bifurcation. Two mutually-conjugate branches of *ghost states* (GSs), with complex propagation constants, are created by the bifurcation, solely in the case of the fractional diffraction. While GSs are not true solutions, direct simulations confirm that their shapes and results of their stability analysis provide a "blueprint" for the evolution of genuine localized modes in the system.

## Introduction

Spontaneous symmetry breaking (SSB) is a ubiquitous phenomenon arising in a wide variety of physical systems [1] which allow the coexistence of symmetric and asymmetric modes with the same propagation constant. SSB of optical solitons was investigated in various settings modeled by the nonlinear Schrödinger equation (NLSE) [2-4] and experimentally observed as well [5,6].

Recently, the work with optical solitons has been expanding in new directions, one of which aims to study nonlinear optical waveguides modeled by NLSE with *PT*-symmetric potentials [7], establishing a connection between conservative and dissipative systems. In such a setting, SSB of optical solitons occurs in a special class of one- and two-dimensional *PT*-symmetric complex potentials [8-14]. On the other hand, optical solitons have been intensively studied in the framework of fractional NLSEs, which address transverse effects governed by fractional-order diffraction, characterized by the respective *Lévy index* (LI) $\alpha$. The linear fractional Schrödinger equation (FSE) originates from fractional-dimensional quantum Hamiltonians, which are produced by the Feynman path integral over Lévy-flight trajectories [15,16]. In terms of classical physics, FSE can be implemented in optical cavities, as proposed by Longhi [17]. Subsequently, the propagation dynamics of optical waves governed by FSE [18,19] was reported, as well as prediction of solitons in various nonlinear settings with fractional diffraction [20-28].

In the framework of the fractional NLSE, SSB of optical solitons has been recently explored under the action of real potentials [29-32]. Our objective here is to consider the symmetry-breaking phenomenology in the framework of the fractional NLSE with *PT*-symmetric potentials, which can be realized in experiment, and offer new possibilities. We report a novel SSB effect for solitons in the fractional NLSE with the Kerr (cubic) or cubic-quintic (CQ) nonlinearity and a *PT*-symmetric potential. It is found that the solitons' *PT*-symmetry is broken with the increase of the power, replacing the solitons by *ghost states* (GSs) with complex propagation constants. They were given this name in Refs. [33-35] because they are not true solutions of the full NLSE, but, nevertheless, the actual evolution of localized states may closely follow the pattern suggested by GSs. We find that, in the present context, they play a similar role: genuine solutions are very close to GSs, as the imaginary part of their propagation constant remains very small, and the instability of GSs against small perturbation is very weak too (unlike the unstable *PT*-symmetric solitons, which are subject to an essentially stronger instability).

## Model

The propagation of paraxial beams under the action of the fractional diffraction and *PT*-symmetric potentials in a nonlinear waveguide is governed by the respective fractional NLSE:

$$2ik_0 \frac{\partial A}{\partial z} - \left(-\frac{\partial^2}{\partial x^2}\right)^{\alpha/2} A + \frac{2k_0^2}{n_0}\left(n(x) - n_0\right)A + \frac{2k_0^2}{n_0} n_{NL} A = 0. \tag{1}$$

Here $A(z,x)$ is the local amplitude of the electromagnetic field with intensity $I \equiv |A|^2$, $k_0 = 2\pi n_0/\lambda$ is the wavenumber, with background refractive index $n_0$ and wavelength $\lambda$, and $(-\partial^2/\partial\xi^2)^{\alpha/2}$ is the fractional-diffraction operator, with LI $\alpha$ belonging to interval $1 < \alpha \leq 2$ [17]. The operator is defined by means of the Fourier transform, with argument $k$: $\hat{F}(-\partial^2/\partial\xi^2)^{\alpha/2}\psi = |k|^\alpha \hat{F}\psi$. Further, the effective potential is $n(x) \equiv n_R(x) + in_I(x)$, where real part $n_R(x)$ represents spatial modulation of the linear refractive index, and imaginary part $n_I(x)$ accounts for distribution of the gain and loss along coordinate $x$, and $n_{NL}(I) = n_2 I + n_4 I^2$ is the cubic-quintic correction to the refractive index. By means of rescaling, $\Psi(\zeta,\xi) \equiv \sqrt{n_4/n_2} A(z,x)$, $\zeta \equiv (k_0 n_2^2/n_0 n_4)z$, $\xi \equiv (2k_0^2 n_2^2/n_0 n_4)^{1/\alpha} x$, and $V(\xi) \equiv n_4(n(x)-n_0)/n_2^2 = V_R(\xi) + iV_I(\xi)$ (it is the *PT*-symmetric complex potential), Eq. (1) is cast in the normalized form,

$$i\frac{\partial \Psi}{\partial \zeta} - \left(-\frac{\partial^2}{\partial \xi^2}\right)^{\alpha/2} \Psi + V(\xi)\Psi + \sigma|\Psi|^2 \Psi + \gamma|\Psi|^4 \Psi = 0, \tag{2}$$

where $\sigma = \pm 1$ and $\gamma \equiv \pm 1$ (or $\gamma = 0$) represent signs of the cubic and quintic terms We consider nonlinearities of two types, *viz.*, the self-focusing Kerr term ($\sigma = 1$, $\gamma = 0$), and the competing CQ combination ($\sigma = 1$, $\gamma = -1$).

Equation (2) amounts to the usual NLSE when LI is $\alpha = 2$. In that case, SSB of solitons is known for a special class of *PT*-symmetric potentials,

$$V(\xi) = g^2(\xi) + hg(\xi) + idg/d\xi, \tag{3}$$

where $g(\xi)$ is a real even function and $h$ an arbitrary real constant. In the combination with the cubic or CQ nonlinearity, potential (3) gives rise to non-*PT*-symmetric solitons with real propagation constants through an SSB bifurcation [8,12].

Setups which realize this propagation regime are based on a Fabry-Perot resonator, with two convex lenses and two phase masks inserted into it. The fractional diffraction is executed in the Fourier representation of the field amplitude [17]. The cubic or CQ nonlinearity is incorporated by placing a piece of a nonlinear optical material between the edge mirror and the closest lens. The *PT* symmetry is realized by adding a balanced pair of gain and loss elements to the resonator, cf. Ref. [36].

In this work, we address the fractional NLSE with potential (3), setting $h = 0$ and adopting

$$g(\xi) = V_0 \left[ \operatorname{sech}\left((\xi + \xi_0)/w_0\right) + \operatorname{sech}\left((\xi - \xi_0)/w_0\right) \right], \qquad (4)$$

with $V_0 = 2$, $\xi_0 = 2$, and $w_0 = 1$. The respective complex potential (3) is displayed in Fig. 1(a).

## Results and Discussions

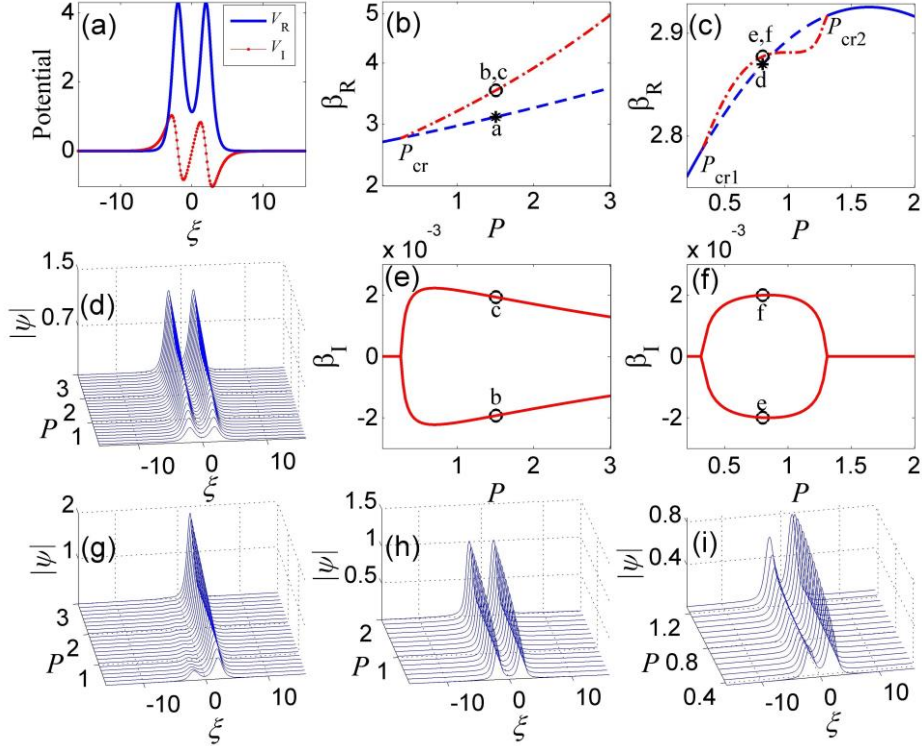

Fig. 1. Symmetry-breaking bifurcations, *PT*-symmetric solitons and GSs produced by Eq. (2) with Lévy index $\alpha = 1.5$ and the self-focusing Kerr ($\sigma = 1$, $\gamma = 0$) or competing CQ (cubic-quintic) nonlinearity ($\sigma = 1$, $\gamma = -1$). (a) The *PT*-symmetric potential defined by Eqs. (3) and (4). (b) The real part of the propagation constant vs. the integral power for stable and unstable *PT*-symmetric solitons (solid and dashed blue lines, respectively) and GSs (ghost states, shown by dashed-dotted red lines) in the model with the Kerr nonlinearity; (c) the same for the CQ model. (e,f) Imaginary parts of the propagation constant of the GSs in (b) and (c), respectively, vs. the power. Points a-f, labeled in panels (b,c,e,f), refer to the solitons and GSs displayed in Fig. 2. Waterfall panels (d) and (g) present, severally, families of *PT*-symmetric solitons and GSs for the Kerr nonlinearity. (h,i) The same for the CQ nonlinearity.

Stationary solutions to Eq. (2) are sought for as $\Psi(\xi,\zeta) = \psi(\xi)\exp(i\beta\zeta)$ with, generally speaking, a complex propagation constant (which corresponds to spatially asymmetric GSs [33-35]), $\beta \equiv \beta_R + i\beta_I$, and function $\psi \equiv \psi_R + i\psi_I$ satisfying the stationary equation, which is written for the practically relevant case of small $\beta_I$:

$$-\left(-\frac{d^2}{d\xi^2}\right)^{\alpha/2}\psi + V(\xi)\psi + \sigma|\psi|^2\psi + \gamma|\psi|^4\psi - \beta\psi = 0. \tag{5}$$

Factors $e^{-2\beta_I\zeta}$ and $e^{-4\beta_I\zeta}$ are omitted in front of the cubic and quintic terms in Eq. (5) for the GSs because $\beta_I$ are actually very small, as shown below. For the same reason, similar factors are omitted in Eqs. (8) and (9) below. Solutions of Eq. (5) can be produced by means of the accelerated imaginary-time evolution method with a fixed power, $P(\beta) = \int_{-\infty}^{+\infty}|\psi(\xi;\beta)|^2 d\xi$ [37].

We first consider stationary solutions of Eq. (2) with LI $\alpha = 1.5$ and the Kerr nonlinearity. The system admits a family of *PT*-symmetric soliton solutions bifurcating from the first discrete linear eigenvalue $\beta \approx 2.71$ of potential $V(\xi)$. With the increase of the soliton power in Fig. 1(b), the solitons undergo destabilization at the critical (bifurcation) point, $P_{cr} \approx 0.28$ ($\beta \approx 2.79$). Beyond the critical point, GSs arise in pairs with complex-conjugate propagation constants. The curve for the real part of the complex propagation constant in Fig. 1(b) represents a new branch bifurcating from the original family of the *PT*-symmetric solitons, while the respective imaginary parts, $\beta_I$, are shown in Fig. 1(e). Note that very small values of $\beta_I$ in the latter plot imply that genuine solutions are indeed very close to the GSs. Figures 1(d) and 1(g) show profiles of the *PT*-symmetric solitons and GSs [the GS in Fig. 1(g) pertains to the upper branch in Fig. 1(e), while the ones associated with the lower branch is $\psi^*(\xi,\beta^*) = \psi(-\xi,\beta)$].

To further analyze the SSB phenomenology, Eq. (5) was also solved with the CQ nonlinearity. In this case, the stable *PT*-symmetric soliton undergoes the destabilization at the first critical point $P_{cr1} \approx 0.34$ ($\beta \approx 2.79$), and then restores its stability at the point of the *inverse SSB bifurcation*, $P_{cr2} \approx 1.28$ ($\beta \approx 2.91$) in Fig. 1(c). Between the two critical points, a pair of GSs with complex-conjugate propagation constants is found. This *bifurcation-loop structure* resembles one which represents the sequence of symmetry-breaking and restoring bifurcations for the dual-core waveguide with the CQ nonlinearity, modeled by a system of linearly-coupled NLSEs [38]. The twisting real parts of the propagation constant in Fig. 1(c) bifurcates from the basic branch of the *PT*-symmetric solitons, crosses it, and eventually merges back into it. Figure 1(h) and 1(i) show profiles of the respective *PT*-symmetric solitons and GSs.

The stability of the *PT*-symmetric solitons and GSs was addressed by adding eigenmodes $u(\xi)$ and $v(\xi)$ of small perturbations, with eigenvalue $\delta \equiv \delta_R + i\delta_I$, to the stationary solution, $\psi(\xi)$:

$$\Psi(\xi,\zeta) = \left[\psi(\xi) + u(\xi)e^{\delta\zeta} + v^*(\xi)e^{\delta^*\zeta}\right]e^{i\beta\zeta}. \tag{6}$$

Substituting it in Eq. (2) and linearizing with respect to the perturbations, we arrive at the linear eigenvalue problem:

$$\begin{pmatrix} L_{11} & L_{12} \\ L_{12}^* & L_{11}^* \end{pmatrix}\begin{pmatrix} u \\ v \end{pmatrix} = \delta\begin{pmatrix} u \\ v \end{pmatrix}, \tag{7}$$

$$L_{11} = i\left[-\left(-\frac{d^2}{d\xi^2}\right)^{\alpha/2} + V - \beta + 2\sigma|\psi|^2 + 3\gamma|\psi|^4\right], \tag{8}$$

$$L_{12} = i\left(\sigma\psi^2 + 2\gamma|\psi|^2\psi^2\right). \tag{9}$$

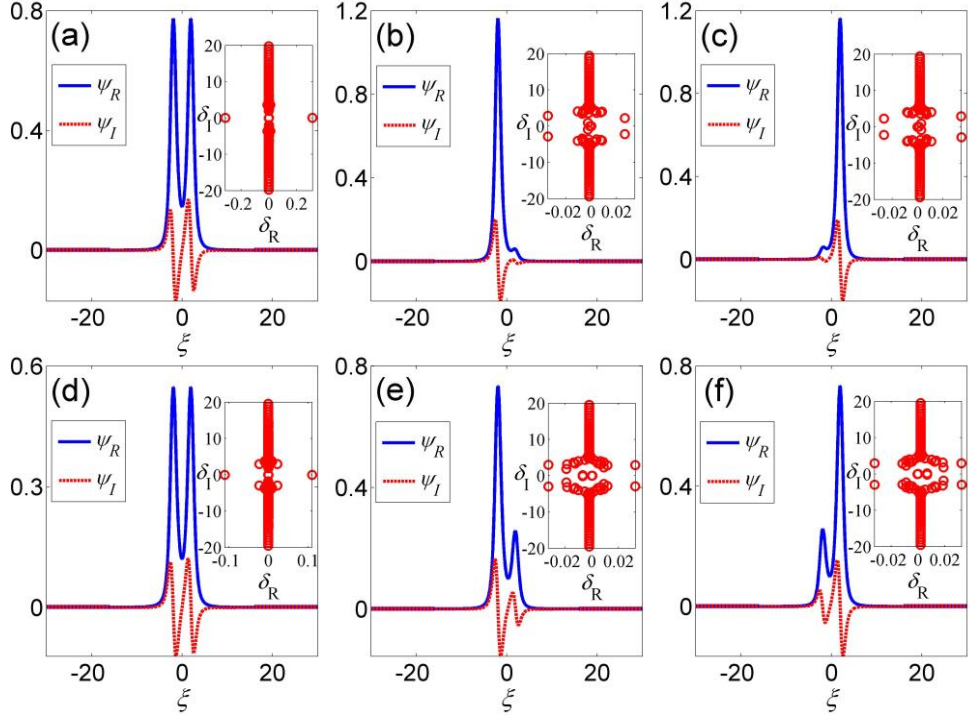

Fig. 2. The top and bottom rows display *PT*-symmetric solitons and GSs (ghost states), along with their linear-stability spectra, for the Kerr (top row) and CQ (bottom row) nonlinearities, respectively, and $\alpha = 1.5$. (a) The *PT*-symmetric soliton with $\beta = 3.12$ corresponding to point "a" in Fig. 1(b). Panels (b) and (c) show, severally, strongly asymmetric GSs with $\beta = 3.56 \mp 0.002i$, corresponding to points "b" and "c" in Figs. 1(b) and 1(e). (d) The soliton with $\beta = 2.87$, corresponding to point "d" in Fig. 1(c). Panels (e) and (f) show, severally, the GSs with $\beta = 2.88 \mp 0.002i$ corresponding to points "e" and "f" in Figs. 1(c) and 1(f).

Equation (7) can be solved by means of the Fourier collocation method [39]. The stationary solution is unstable if there is at least one eigenvalue with $\delta_R > 0$. Similar to the fact that the GSs are not true solutions of the NLSE, but rather "blueprints" for their evolution, the prediction of formal stability or weak instability of the GS determines robustness of genuine solutions.

As generic examples, a *PT*-symmetric soliton and GSs, along with their stability spectra, are presented in Figs. 2(a), (b) and (c) for the Kerr nonlinearity, with power $P = 1.5$, and in 2(d), (e) and (f) for CQ, with $P = 0.8$. Note spatial asymmetry of GS profiles.

The stability analysis demonstrates that the *PT*-symmetric solitons are stable where GSs do not exist, and unstable in the domain of the broken symmetry, while all GSs are unstable. However, the growth rates of the GS, $\delta_R$, are smaller than the growth rate of *PT*-symmetric solitons in Figs. 2 by a factor of $\sim 5-10$, which implies that genuine solutions, following the GS "blueprints", may be quasi-stable, as confirmed by direct simulations below.

Next, we then address the effect of LI $\alpha$ on the GSs. To this end, numerically found dependences of the real and imaginary parts of their propagation constant on the power are displayed in Fig. 3 for $\alpha$ ranging from 1.1 to 2 with a step of 0.1. It is seen in panels 3(b) and (d) that the GSs, with $\beta_I \neq 0$, exist *solely* for the fractional dimension. They carry over into asymmetric solitons with a real propagation constant at $\alpha = 2$, which are the non-*PT*-symmetric solitons from Refs. [8,9].

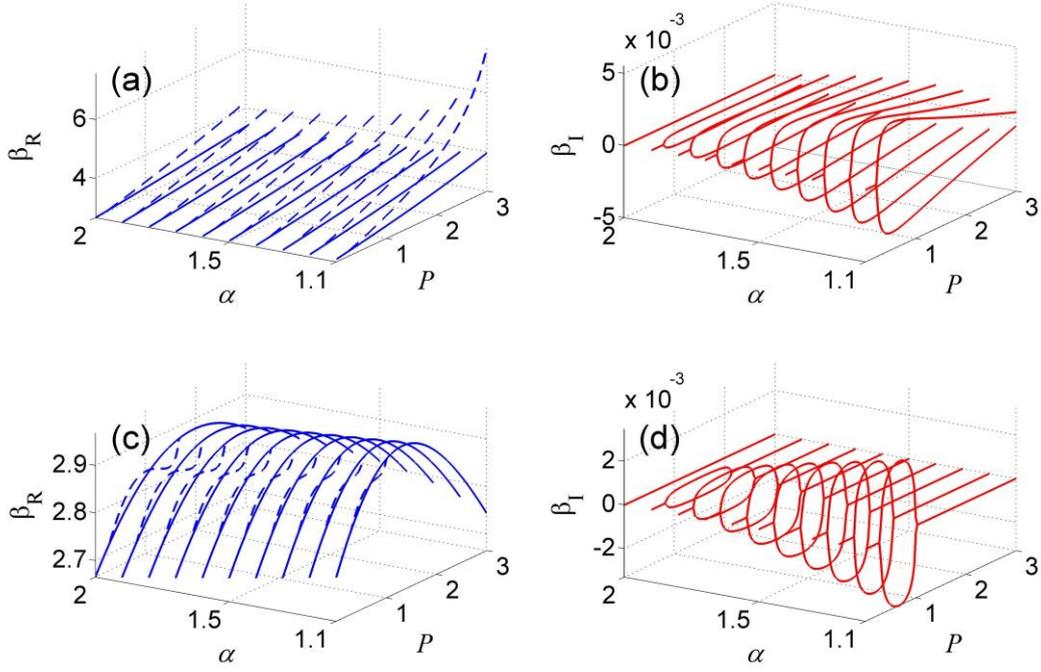

Fig. 3. Dependences of the real and imaginary parts of the propagation constant of the GSs on Lévy index $\alpha$ and power $P$: (a,b) the Kerr nonlinearity; (c,d) CQ.

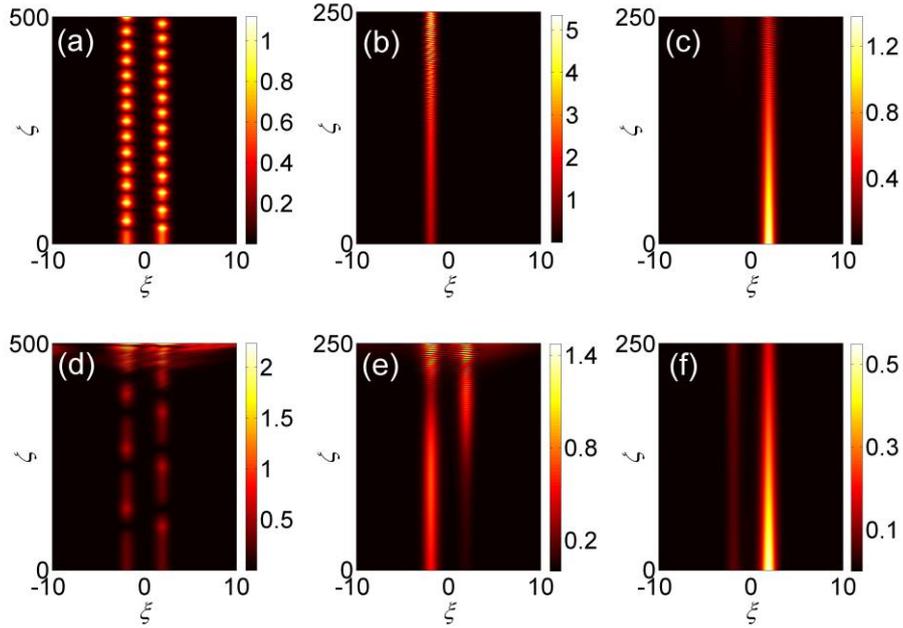

Fig. 4. Dynamics of *PT*-symmetric solitons and GSs from Fig. 2 (for $\alpha = 1.5$). (a,d): The unstable evolution of *PT*-symmetric solitons from Figs. 2(a,d) over the propagation distance $\zeta > 20$ Rayleigh lengths, $\zeta_R$. Panels (b,c) and (e,f) display the evolution of the GSs from 2(b,c) and (e,f), respectively, for $\zeta > 10\zeta_R$.

The perturbed dynamics of the stationary states in the present model was explored by dint of direct simulations of Eq. (2). In the SSB domain, the *PT*-symmetric solitons develop conspicuous instability (coupled to intrinsic oscillations of the modes) even without addition of initial perturbations, as seen in Figs. 4(a,d). On the other hand, the GSs survive as quasi-stable states, if perturbed by 1% random noise in Figs. 4(b,c,e,f) at the early stage of the evolution ($\zeta < 50$), then developing oscillatory perturbations which, however, do not completely destroy the GSs at $50 < \zeta < 100$. Later, some GSs suffer complete destruction [in 4(b,e)], while others survive, as perturbed modes, over longer propagation distances, such as ones in Figs. 4(c,f). At $\zeta > 250$, the above-mentioned effect of $\beta_I$ on Eqs. (5) cannot be neglected.

**Conclusions**

In summary, we have reported a scenario of SSB (spontaneous symmetry breaking) of solitons in the fractional NLSE with the cubic or CQ (cubic-quintic) nonlinearity and *PT*-symmetric potential. Solely in the case of the fractional diffraction, the SSB creates spatially asymmetric GSs ("ghost states") with a complex propagation constant. Although not being true solutions of the underlying NLSE, they provide "blueprints" for the quasi-stable evolution of genuine solutions. The setting can be realized in nonlinear optical cavities.

A relevant direction for extension of the analysis may be the consideration of a two-dimensional version of the system.

**Funding**

National Natural Science Foundation of China (NNSFC) (11805141); Applied Basic Research Program of Shanxi Province (201901D211424); Scientific and Technological Innovation Programs of Higher Education Institutions in Shanxi (STIP) (2019L0782); "1331 Project" Key Innovative Research Team of Taiyuan Normal University (I0190364); Israel Science Foundation, grant No. 1286/17.